\documentclass{article}

\usepackage[final]{nips_2018}

\usepackage[utf8]{inputenc} 
\usepackage[T1]{fontenc}    
\usepackage{hyperref}       
\usepackage{url}            
\usepackage{booktabs}       
\usepackage{amsfonts}       
\usepackage{nicefrac}       
\usepackage{microtype}      
\usepackage{graphicx}
\usepackage[linesnumbered,ruled]{algorithm2e}
\usepackage{algorithmic}

\usepackage{graphicx}
\usepackage{subcaption}
\usepackage{floatrow}
\title{Reproducing AmbientGAN: Generative models from lossy measurements }

\author{
  Mehdi Ahmadi\\
  Polytechnique Montreal\\
  \texttt{mehdi.ahmadi@polymtl.ca} \\
   \And
   Timothy Nest \\
   University de Montreal\\
  \texttt{timothy.nest@mail.mcgill.ca} \\
  \AND
   Mostafa Abdelnaim \\
  University de Montreal\\
  \texttt{mostafa.abdelnaim@umontreal.ca} \\
   \And
  Thanh-Dung Le \\
  Institut national de la recherche scientifique \\
  \texttt{dung.le@emt.inrs.ca} \\
}

\begin{document}

\maketitle

\section{Paper Summary}
\subsection{Motivation}
In recent years, Generative Adversarial Networks (GANs) have shown substantial progress in modeling complex distributions of data. These networks have received tremendous attention since they can generate implicit probabilistic models that produce realistic data using a stochastic procedure. While such models have proven highly effective in diverse scenarios, they require a large set of fully-observed training samples. In many applications access to such samples are difficult or even impractical and only noisy or partial observations of the desired distribution is available. Recent research has tried to address the problem of incompletely observed samples to recover the distribution of the data. \citep{zhu2017unpaired} and \citep{yeh2016semantic} proposed methods to solve ill-posed inverse problem using cycle-consistency and latent-space mappings in adversarial networks, respectively. \citep{bora2017compressed} and \citep{kabkab2018task} have applied similar adversarial approaches to the problem of compressed sensing. 
In this work, we focus on a new variant of GAN models called AmbientGAN, which incorporates a measurement process (e.g. adding noise, data removal and projection) into the GAN training. While in the standard GAN, the discriminator distinguishes a generated image from a real image, in AmbientGAN model the discriminator has to separate a real measurement from a simulated measurement of a generated image. The results shown by \citep{bora2018ambientgan} are quite promising for the problem of incomplete data, and have potentially important implications for generative approaches to compressed sensing and ill-posed problems. \\

\subsection{Proposed approach}
\begin{figure}
  \includegraphics[width=\linewidth]{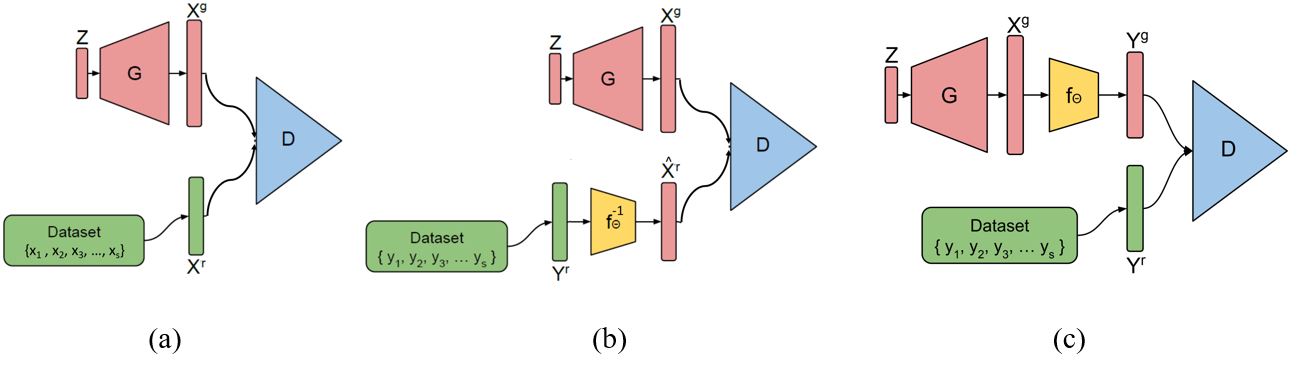}
   \caption{(a) Standard GAN (b) Baseline model (c) AmbientGAN }
  \label{fig:fig1}
\end{figure}

The original Generative Adversarial Network proposed by \citep{goodfellow2014generative} tries to map an easy-to-sample distribution (e.g. a low-dimensional Gaussian distribution) to a high-dimensional domain of interest such as images. As shown in Fig. \ref{fig:fig1}(a), the standard GAN is composed of a generator that attempts to produce a realistic image, and a discriminator unit that determines which images are real and which have been produced by the generator. In standard GAN, the dataset contains fully-observed samples without any measurement process such as noise or projection. One approach to deal with the dataset from incomplete or measurement distribution is to utilize the inverse of the measurement function $f_\theta$ and obtain the inverted samples (i.e. unmeasured samples) for training process. As shown in Fig. \ref{fig:fig1}(b),  the  estimated inverse samples for a given measurement are utilized to learn a generative model in order to approximate the distribution of real data $p_x^r$. This model is called a baseline model in the original AmbientGAN paper, since some of the measurement models are not invertible, the authors utilize approximation for the invers function. 

The AmbientGAN model adapts the original GAN configuration in an efficient way to handle cases in which the dataset consists of noisy or incomplete samples. The idea behind the AmbientGAN model is to apply the same measurement process to the output of the generator. As shown in Fig. \ref{fig:fig1}(c), the discriminator in the AmbientGAN model has to differentiate the measurement of a generated image from the real measurement. Since the distribution of measured images uniquely determines the distribution of original images, training the AmbientGAN model results in a generator that produces images similar to real images without measurement.

In Fig. \ref{fig:fig1}(c), let $Z\in R^k$ be the latent variable vector with the distribution $p_z$, which can be a Gaussian or uniform distribution. Let generator $G: R^k \to R^n$ produces generated samples $X^g$ with the distribution $p_x^g$. The measurement function $f_\theta:R^n \to R^m$ is parametrized by $\theta$ which has the distribution $p_\theta$ and outputs $Y^g=f_\theta(X^g)$ with the distribution $p_y^g$. Since the desired samples $X\sim p_x^r$ are not available, we are given a set of IID samples from measurement distribution $Y\sim p_y^r$ as the training dataset. As in the standard GAN, the objective function for the AmbientGAN model is a min-max scenario as described in:
\begin{equation}
\min_{G}\max_{D} \mathbb{E}_{Y^r \sim p_y^r}\Big[q(D(Y^r))\Big] +\mathbb{E}_{Z \sim p_z,\Theta \sim p_\theta}\Big[q
(1-D(f_\Theta(G(Z))))\Big]
\end{equation}

where $q(.)$ is the quality function (e.g. $q(x)=log(x)$ for the standard GAN). The goal of this objective function is to learn a generator $G$ such that $p^g_x$ is close to $p_x^r$ (since if $X\sim p_x^r$ and $\Theta\sim p_\theta$, then $Y=f_\Theta(X) \sim p_y^r$).\\

\textbf{Assumptions}: One assumption is that the measurement function $f_\theta$ is known and it is easy to sample $\Theta\sim p_\theta$. In addition, $f_\theta$ is required to be a differentiable function with respect to its inputs for all $\theta$, in order to have an end-to-end differentiable model for training. Another important assumption is that for a given observed measurement distribution $p_y^r$, there is a unique true underlying distribution $p_x^r$. As the result, if the discriminator is optimal, i.e. $D(.) = \frac{p_y^r(.)}{p_y^r(.) + p_y^g(.)}$, then a generator $G$ is optimal iff $p^g_x=p^r_x$. In other words, if the distribution of the measured generated images are close to the measured real images, the distribution of the generated images without measurement will be close to the distribution of the real images. The aforementioned assumption and its results Lemma is valid for certain measurement models, which are introduced in the original paper. 

\section{Experiments}
\subsection{Motivation for the experiments}
Since the AmbientGAN framework aims to model the distribution of the true data based on the available noisy and incomplete training samples, we are motivated to compare its effectiveness relative to the baseline model. As mentioned earlier, in the baseline model the training samples are first cleaned up using the inverse of the measurement function (or its approximation) and then they are delivered to the discriminator. We also favor to know the capability of the AmbientGAN model to recover the true underlying distribution in the presence of different measurement models. The measurement models used in the paper are listed as follows:

\begin{itemize}
   \item Block-Pixels: where each pixel is set to zero with probability $p$.
  \item Convolve-Noise: where images are convolved with a Gaussian kernel $k$, and noise is added.
  \item Keep-Patch: where pixels outside of a randomly chosen $k\times k$ patch are set to zero.
  \item Extract-Patch: where pixels within a randomly chosen $k\times k$ are extracted.
  \item Pad-Rotate-Project: where the padded image is  rotated at a random angle about the center.
  \item Pad-Rotate-Project-$\theta$: where the chosen angle is also included in the measurement process.
  \item Gaussian-Projection: where the image is projected onto a random Gaussian vector.
 \end{itemize}
\subsection{Reproducing the main results}
In order to verify the results of the AmbientGAN model, we performed a set of experiments similar to the ones in the paper. For implementation, we took inspiration from two GitHub repositories $\footnote{https://github.com/shinseung428/ambientGAN\_TF}$,$\footnote{https://github.com/AshishBora/ambient-gan}$  and modified the codes according to requirements for each experiment. The pseudo code to perform AmbientGAN training is illustrated in the Algorithm \ref{alg:1}. In the following algorithm, after sampling from $Z\sim p_g(z)$, the step of applying the measurement function $f_\theta$ to the generated samples is added to the standard GAN training.

\begin{algorithm}
  \begin{algorithmic}[1]
    \FOR{number of training iterations}
      \FOR{$k \in \{1,\dots,K\}$}
        \STATE Sample mini-batch of m noise samples $\{z_1,..,z_m\}$ form noise prior $p_g(z)$.
        \STATE Sample mini-batch of m measured samples $\{y_1^g,..,y_m^g\}$ form measurement function distribution $p_y^g$.
        \STATE Sample mini-batch of m examples $\{y_1^r,..,y_m^r\}$ form data generating distribution $p_y^r$.
        \STATE Update the discriminator by ascending its stochastic gradient:$$\nabla_{\theta_d}\frac{1}{m}\sum_{i=1}^{m}\Big[q(D(y_i^r))+q\Big(1-D(f_\theta(G(z_i)))\Big)\Big]$$
     \ENDFOR
    \STATE Sample mini-batch of m noise samples $\{z^{(1)},..,z^{(m)}\}$ form noise prior $p_g(z)$.
    \STATE Sample mini-batch of m measured samples $\{y_1^g,..,y_m^g\}$ form measurement function distribution $p_y^g$.
    \STATE Update the generator by descending its stochastic gradient:$$\nabla_{\theta_d}\frac{1}{m}\sum_{i=1}^{m}q\Big(1-D(f_\theta(G(z_i)))\Big)$$
    \ENDFOR
  \end{algorithmic}
   \caption{AmbientGAN Algorithm}
   \label{alg:1}
\end{algorithm}

We repeated the experiments in the paper on several datasets. For the celebA dataset, Block-Pixels, Convolve-Noise, Keep-Patch and Pad-Rotate-Project measurements were evaluated. For the CIFAR-10 dataset, Block-Pixels measurement experiment was experimented and for MNIST dataset Pad-Rotate-Project and Pad-Rotate-Project-$\theta$ were applied. To run all mentioned experiments, we executed the code using Tensorflow library on a TITANX GPU, a GTX 1080 Ti GPU, a Tesla k80 GPU from Colab service of Google,  and k20 GPUs from Guillimin compute cluster. Fig. 2 and 3 show the results of applying different measurements on celebA (with 35000 training iterations), CIFAR-10 datasets (with 25000 training iterations), respectively. The results for MNIST dataset (with 70000 training iterations) are presented in the next section.

\begin{figure}
    \centering
    \begin{subfigure}[b]{1\textwidth}
        \includegraphics[width=\textwidth]{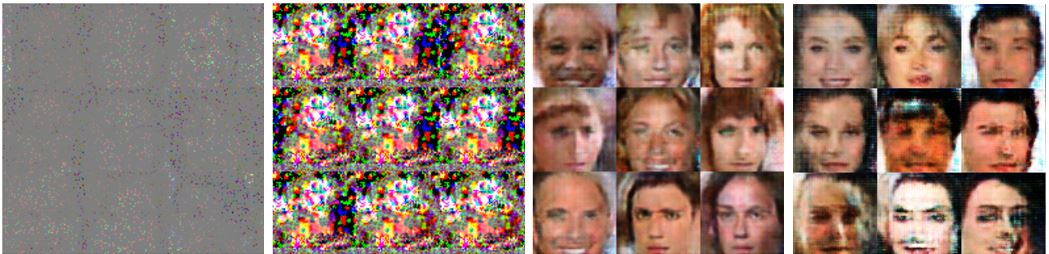}
        \caption{}
        \label{fig:fig2a}
    \end{subfigure}
    ~ 
    \begin{subfigure}[b]{1\textwidth}
        \includegraphics[width=\textwidth]{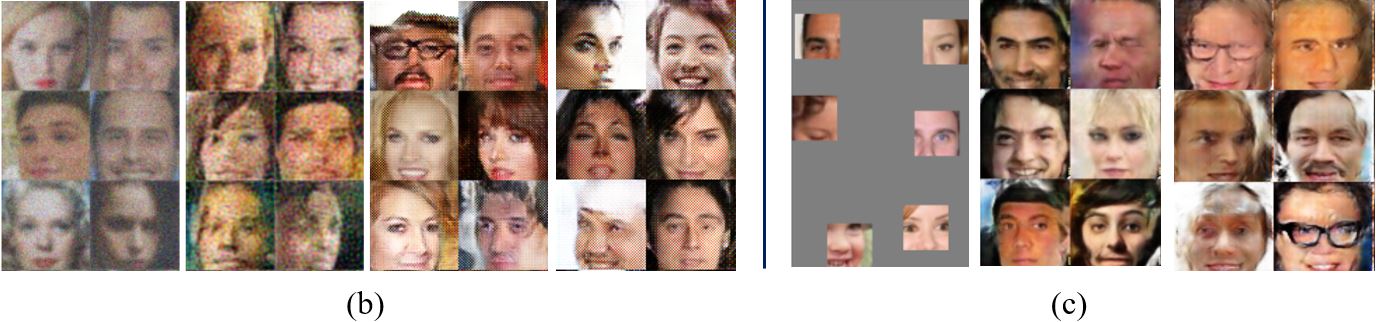}
        \label{fig:fig2c}
    \end{subfigure}
    \begin{subfigure}[b]{1\textwidth}
        \includegraphics[width=\textwidth]{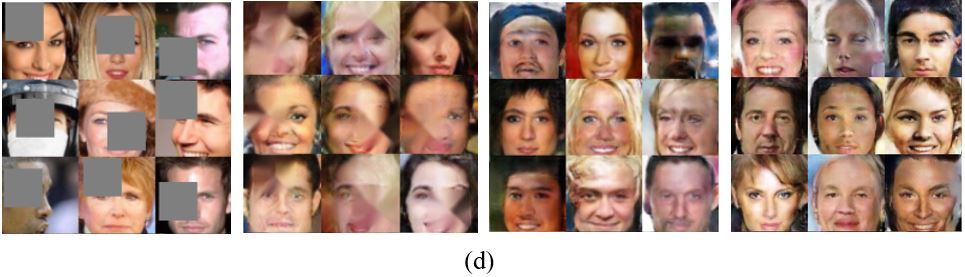}
        \label{fig:fig2d}
    \end{subfigure}
    
    ~ 
    \caption{Results on celebA with: (a) Block-Pixels ($p$=0.95), (b) Convolve-Noise,  (c) Keep-Patch and (d) Block-Patch.  Left: samples from lossy measurement, middle left: baseline model, middle right: AmbientGAN from the original paper, right: AmbientGAN reproduced in this work. For the Keep-Patch measurement, there is no baseline model.}
    \label{fig:fig22}
\end{figure}

As shown in Fig. \ref{fig:fig22}, the AmbientGAN model is powerful enough to produce faces with acceptable visual quality, even though most of the pixels in the training samples are heavily degraded with Block-Pixels measurement ($p$=0.95). The baseline model, on the other hand, does not show good results since the inverse of measurement function cannot perfectly clean the measured samples when dealing with severe measurement process. The images in the right side are our reproducing results that confirm the AmbientGAN results in the original paper.

\begin{figure}
  \includegraphics[width=\linewidth]{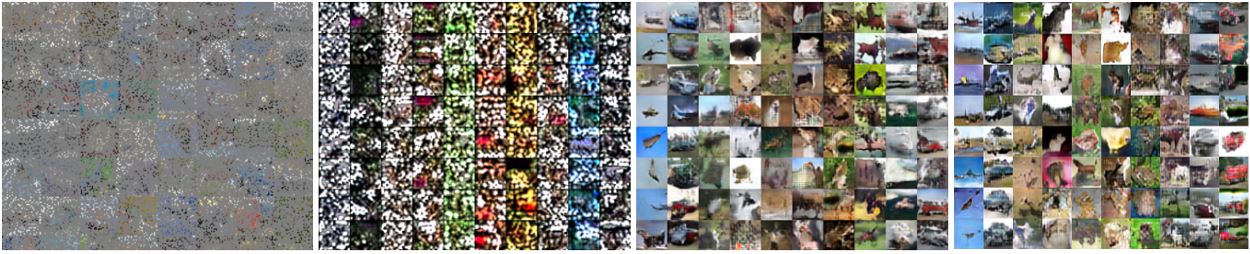}
  \caption{Results on CIFAR-10 with block pixel ($p$=0.8). Left: samples from lossy measurement, middle left: baseline, middle right: AmbientGAN from original paper, right: Ambient GAN reproduced in this work.}
  \label{fig:fig33}
\end{figure}

\begin{figure}
  \centering
  \begin{subfigure}[b]{0.45\linewidth}
    \includegraphics[width=\linewidth]{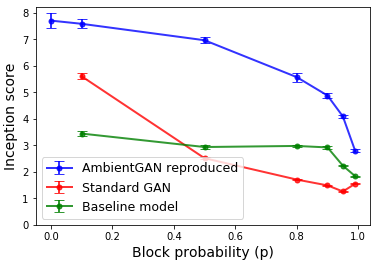}
    \label{fig:Fig4_Inception_scores_a}
  \end{subfigure}
  \begin{subfigure}[b]{0.45\linewidth}
    \includegraphics[width=\linewidth]{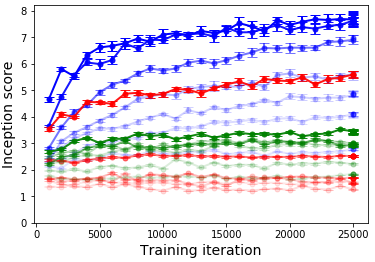}
    \label{fig:Fig4_Inception_scores_b}
  \end{subfigure}
  \caption{Reproducing quantitative results for CIFAR-10 datasets. (Left) Inception score vs blocking probability $p$. (Right) Inception score vs training iteration with darkness proportional to $(1-p)$.}
  \label{fig:fig4}
\end{figure}

The AmbientGAN also shows promising results for other datasets. As shown in Fig. \ref{fig:fig33}, for CIFAR-10 dataset in the presence of Block-Pixels measurement with with $p$=0.8, the AmbientGAN can produce samples  similar to CIFAR-10 images with high quality while the baseline model fails to produce meaningful images. In addition to qualitative results, we also reproduced the experiments for the quantitative results using an inception model$\footnote{http://download.tensorflow.org/models/image/ imagenet/inception-2015-12-05.tgz}$ trained on ImageNet dataset as shown in Fig. \ref{fig:fig4}.

As shown in Fig. \ref{fig:fig4}, by increasing the blocking probability, the inception score of AmbientGAN model does not degrade quickly while for the baseline model and the standard GAN model which is trained with measured training samples (i.e. it ignores any measurement process), the inception score dramatically degrades. In addition, by increasing the training iteration, the inception score of the AmbientGAN model constantly improves and it is higher than other models with the same $p$.

\section{Discussion}
In this section, we first explore different aspects of the experiments and discuss the missing information in the paper. Then, we conclude the report by illustrating how the experiments are aligned with the proposed analysis.

\textbf{Choosing Hyperparameters:} The AmbientGAN structure consists of a generator and discriminator similar to the standard GAN model with an additional unit of measurement function $f_\theta$. The authors choose the same implementation and hyper parameters as described in \citep{radford2015unsupervised} and \citep{arjovsky2017wasserstein}, for the units which are common with the standard GAN. Therefore, the only unit that introduces new parameters in AmbientGAN is the measurement function. For each experiment, we change the values of the parameters to see the effect of measurement process on the AmbientGAN. Table \ref{table:table2} shows the selected hyper parameters and measurement   parameters.
\begin{table}[!ht]
\resizebox{1\textwidth}{!}{\begin{minipage}{\textwidth}
\centering
\scalebox{0.9}{
\begin{tabular}{| c | c | c | c |}
    \hline
    \multicolumn{2}{|c|}{\textbf{Generator \& Discriminator units}}&\multicolumn{2}{|c|}{\textbf{Measurement units}}\\ \hline
    Hyper Parameters & Value & Parameters & Value \\ \hline
    Number of Epochs & 25 & Block pixel probability & [0.5, 0.8, 0.9, 0.95, 0.98] \\ \hline
    Optimization method & Adam & Patch size & 32x32 \\ \hline
    Learning rate & 0.0002 & Number of angles & 1 \\ \hline
    Momentum & 0.5 & Noise ($\mu$ and $\sigma$ ) & [(0, 0.1), (0, 0.2)] \\ \hline
    Batch size & 64 & blur Kernel size & 5x5 \\ \hline
    \end{tabular}
\par}
\caption{The AmbientGAN hyper parameters and measurement parameters}
\label{table:table2}
\end{minipage} }
\end{table}

\textbf{Failure cases and missing information:}
The AmbientGAN describes an approach to deal with noisy and incomplete training samples corrupted by some common measurement models such as noise and projection. However, the simulation results for Padding-Rotate-Project-$\theta$ does not demonstrate acceptable results for CelebA. As shown in Fig. 5, AmbientGAN only generates a general outline of the face without clear visualization of the elements inside the face. This implies that the AmbientGAN has difficulty to learn the complex distributions thorough 1D projection.
\begin{figure}
  \centering
  \begin{subfigure}[b]{0.3\linewidth}
    \includegraphics[width=\linewidth]{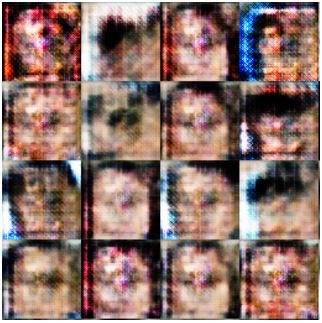}
    \caption{}
    \label{fig:Fig5_a}
  \end{subfigure}
  \begin{subfigure}[b]{0.3\linewidth}
    \includegraphics[width=\linewidth]{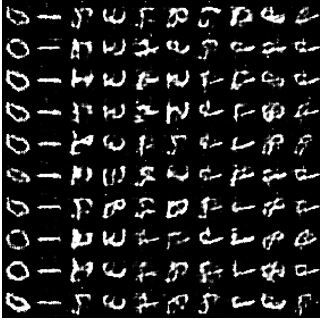}
    \caption{}
    \label{fig:Fig5_b}
  \end{subfigure}
  \begin{subfigure}[b]{0.3\linewidth}
    \includegraphics[width=\linewidth]{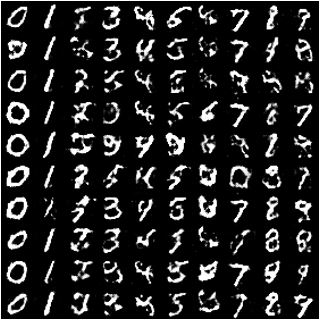}
    \caption{}
    \label{fig:Fig5_c}
  \end{subfigure}
  \caption{Generated samples by AmbientGAN: (a) CelebA dataset with 1D projection, (b) MNIST dataset with Pad-Rotate-Project-$\theta$, (c) MNSIT dataset with Pad-Rotate-Project.}
  \label{fig:Fig5}
\end{figure}

As shown in Fig. 5 (b) and (c), AmbientGAN can generate the digits with correct orientation, if the value of the theta is provided. However, even for the MNIST dataset which has simpler distribution compared with CelebA dataset, the AmbientGAN still has problems in producing readable digits. The paper also introduce the Gaussian projection as a measurement model but there is no discussion or simulation results about training AmbientGAN in the presence of this measurement. Reconstruction images from projection is a common technique in applications such as MRI and the success of AmbientGAN in this area is not fully understood from the original paper.

\begin{figure}
  \centering
  \begin{subfigure}[b]{0.302\linewidth}
    \includegraphics[width=\linewidth]{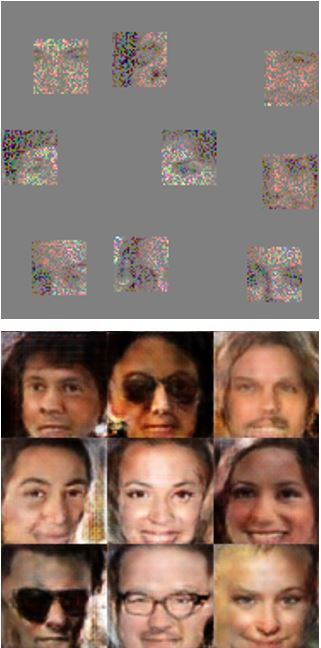}
    \caption{}
    \label{fig:Fig6_a}
  \end{subfigure}
  \begin{subfigure}[b]{0.3\linewidth}
    \includegraphics[width=\linewidth]{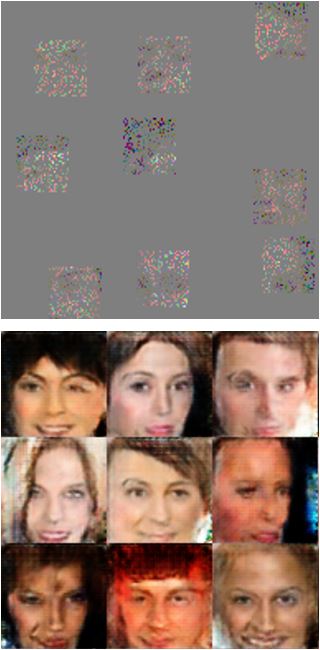}
    \caption{}
    \label{fig:Fig6_b}
  \end{subfigure}
  \begin{subfigure}[b]{0.3\linewidth}
    \includegraphics[width=\linewidth]{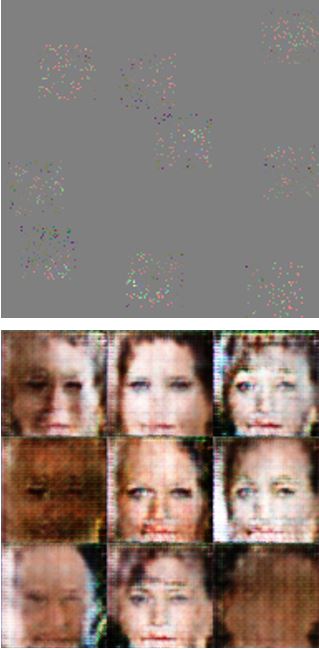}
    \caption{}
    \label{fig:Fig6_c}
  \end{subfigure}
  \caption{Combined measurements Keep-Patch and Block-Pixels probability: (a) $p$=0.5, (b) $p$=0.85, (c) $p$=0.95}
  \label{fig:Fig6}
\end{figure}

Another issue with AmbientGAN paper is missing the experiments about measurements in which the location information is lost. For example, there is no simulation results about Extract patch, to show the behavior of Ambient GAN in generating images when the training images are included some parts of faces without being in their natural locations.

\textbf{Extra Experiments:} The paper evaluates the AmbientGAN model for each measurement process separately.  However, there might be more than one source of corruption on image in practice. Therefore, we applied both keep patch and block pixels, as two sources of measurements. Among different patch measurements, keep patch removes most part of the samples and its combination with block pixels introduce even severe corruptions on samples. Fig. 6, shows the simulation results for AmbientGAN in the presence of the combined measurements.

To perform the combined noise experiment, the bigger the block pixel probability is applied, the longer the training time is required so that the model can converge. As shown in Fig. 6, moreover, we can see the results for combining keep patch and block pixel with probability 0.5, 0.85 and 0.95, respectively. For the first two cases, the AmbientGAN can still effectively generated samples. Unfortunately, in case of probability of 0.95, AmbientGAN cannot produce a high visualization result, although, AmbientGAN can generate a well-observed image when we solely applied block pixel probability of 0.95 as shown in previous experiment.

As shown in Fig. 6 even with increasing the amount of distortion in the image, therefore, the AmbientGAN still can generate good faces.
one interesting thing is that in almost all the experiments in CelebA, we have the location information. but we do not know what happens when the location information is missed. we are wondering why the authors did not report the results for extract patch since the location information is not available in that measurement. For pad-rotate, maybe one of the reasons  bad results comes from the missing the location.

In the AmbientGAN paper, the authors tried a variety of measurement processes on different datasets to demonstrate that the proposed AmbientGAN generates better results than the baseline model. They reported some severe measurement process in which the training samples are degraded heavily but the AmbientGAN is still able to generate image perfectly. The author also provided theoretical analysis and proved the strong assumption of the paper that the distribution of measured images uniquely determines the distribution of original images. In this section, we explore different aspects of the experiments and discuss the missing information in the paper.
\bibliography{iclr2017_conference}
\bibliographystyle{iclr2017_conference}
\end{document}